\documentclass[a4paper,12pt]{article}
\usepackage[utf8]{inputenc}
\usepackage{amsmath,amsbsy}
\usepackage{natbib}
\usepackage{graphicx}
\usepackage[margin=1in]{geometry}

\newcommand{\vv}{\ensuremath{\mathbf{v}}}
\newcommand{\xx}{\ensuremath{\mathbf{x}}}

\newcommand{\M}{\ensuremath{\mathcal{M}}}
\newcommand{\T}{\ensuremath{\mathcal{T}}}
\newcommand{\tB}{\ensuremath{t_{B}}}
\newcommand{\tE}{\ensuremath{t_{E}}}
\newcommand{\XX}{\ensuremath{{\mathcal{X}}}}
\newcommand{\YY}{\ensuremath{{\mathcal{Z}}}}
\newcommand{\VV}{\ensuremath{{\mathcal{V}}}}
\newcommand{\ones}{\ensuremath{\mathbf{1}}}
\newcommand{\transpose}{\textsf{T}}
\newcommand{\dist}[1]{\text{#1}}
\newcommand{\errorSD}{\varsigma}
\newcommand{\tallyprop}[1]{\vspace*{1ex}\noindent\textbf{#1~}}

\title{\textbf{Bayesian inference for velocity-jump models for movement}}
\author{Paul G.\ Blackwell\\
School of Mathematical and Physical Sciences\\
University of Sheffield\\
Sheffield S3 7RH, U.K.\\
Email: p.blackwell@sheffield.ac.uk}
\date{}

\begin{document}

\maketitle

\section*{Abstract}
The velocity-jump model is a specific type of piecewise deterministic Markov process in which an individual's velocity is constant except at times that form the events of some point process. It represents an interpretable continuous-time version of the discrete-time `step and turn' models widely used in analysing wildlife telemetry. In this paper, I derive a reversible jump Markov chain Monte Carlo algorithm to carry out exact Bayesian inference for velocity-jump models by reconstructing the trajectories between observations, and illustrate its use in analysing real and simulated telemetry data. The method uses a proposal distribution for updating velocities that is constructed by approximating the movement model with a multivariate normal distribution and then conditioning that distribution on the data. The velocity-jump models considered can incorporate measurement error and Markov dependence between successive velocities.

\noindent
Keywords: animal movement; 
piecewise deterministic Markov process; 
reversible jump Markov chain Monte Carlo; telemetry; velocity-jump process

\section{Introduction}

The trajectories of individual animals are often conceptualised as continuous, piecewise linear paths, going back at least to the run-and-tumble models of \citet{Othmer}. 
In more recent applied work, a common practice 
is to take an animal's trajectory based on regularly observed locations, and model it in terms of the distance travelled over each interval and the turning angle between successive segments of this supposed path. Such step-and-turn models in discrete time are widely used \citep{toby}, 
but suffer the usual theoretical and practical limitations of the discrete-time modelling of a continuous-time process. 
A more specific drawback is that, since turns are generally assumed to take place at all observed time points, they are imposed rather than inferred, and so it does not make sense to interpret their existence literally as representing meaningful changes in an animal's path, tempting though that might be.

One continuous-time approach to capturing some of the same intuition about movement is to use a diffusion model expressed in terms of bearing and speed. Such a continuous-time step-and-turn model is defined and explored by \citet{AlisonBAYSM} and \citet{AlisonJABES}, and discussed further in Section \ref{parton}. Another approach is to maintain the straight-line movement, but allow the times of changes in velocity to 
form a point process, regardless of the timing of observations. That is the approach taken by \citet{RuizSuarez}; however, they place quite strong restrictions on some elements of the model, and they use Approximate Bayesian Computation for their inference, which limits the information that can be recovered about actual trajectories. The main aim of the current paper is to develop a fully Bayesian inference method for a more general version of their model. I'll refer to this model as a 
velocity-jump model, since `step and turn' is ambiguous in this context.
This velocity-jump model is a special case of the piecewise deterministic Markov processes (PDMPs)
of \citet{Davis1993}, where the state space of the process represents location and velocity, and the deterministic part has velocity unchanging and location changing according to the current velocity.

The model is defined fully in Section \ref{model}, and the inference methodology is described in Section \ref{infer} and applied in Section \ref{apply}. 
Other aspects are discussed in Section \ref{discuss}.

\section{The Velocity-Jump Model}
\label{model}

The velocity-jump movement model is defined as follows in two dimensions; higher-dimensional extensions are straightforward.
Let $t_1,t_2,\ldots$, representing the times of events at which velocity changes, form a Poisson process of rate $\lambda$ on $[0,\infty)$.
Let $\theta_1,\theta_2,\ldots$ represent i.i.d.\ turning angles from some circular distribution typically centred at 0; throughout the paper I will take 
$\theta_i \sim \dist{von Mises}(0, \kappa)$
where $\kappa$ is a concentration parameter, though essentially any circular distribution with non-arithmetic support could be used.
Let $s_0,s_1,s_2,\ldots$ represent speeds, with $s_i$ applying between $t_i$ and $t_{i+1}$,
where $t_0$ represents the time of the beginning of the process, or the initial observation; often for convenience $t_0\stackrel{\Delta}{=}0$. Speeds are taken to be identically distributed 
and to either be independent or form a first-order Markov process; in either case they are assumed to be independent of the turns.

Given an initial bearing $\psi_0$, the animal's bearings over time are given by 
\[
\psi_{i+1} = \psi_{i} + \theta_{i+1} , ~~i\geq 1.
\]
If the initial bearing $\psi_0$ has a uniform distribution on $[0,2\pi)$, then the marginal distribution for each $\psi_i, i\geq 1$ is also uniform on $[0,2\pi)$, 
regardless of the turning angle distribution. 
Then, given an initial location $\xx_0$, the animal's locations at the event times are given by 
\[
\xx_{i+1} = \xx_{i} + (t_{i+1}-t_i)s_i (\cos(\psi_{i}),\sin(\psi_{i}))^\transpose, ~~i\geq 1.
\]
Its locations between events, including at observation times, are given by 
\[
\xx(t) = \xx_i + (t-t_i) s_i (\cos(\psi_{i}),\sin(\psi_{i}))^\transpose, ~~t\in[t_i,t_{i+1}].
\]

\subsection{Gaussian velocities}

If the speeds follow independent $\dist{Rayleigh}(\sigma)$ distributions, and the marginal distribution for the bearing is uniform on $[0,2\pi)$ as above, then the marginal distributions for the velocities $\vv_i = s_i (\cos(\psi_{i}),\sin(\psi_{i}))^\transpose$ will be circular bivariate Gaussian $\dist{N}((0,0)^\transpose,\sigma^2I_2)$. The same applies in the case of first-order Markov dependent $\dist{Rayleigh}(\sigma)$ speeds---see Section \ref{rice} for details---provided the initial speed is also $\dist{Rayleigh}(\sigma)$. Bivariate Gaussian velocities are often assumed in diffusion models of animal movement \citep{johnson2008,TheoVelocity}, and are used in the examples in \S\ref{apply}.

Note that even when velocities have Gaussian marginal distributions, their joint distributions are \emph{not} Gaussian; instead, their dependence has a more realistic form inherited from the turning angle mechanism, and from the dependence between speeds if applicable. 

Since the Rayleigh distribution is a special case (with shape parameter 2) of the Weibull distribution, and the speed distribution in discrete-time models is often Weibull, it is natural to allow speeds to come from a more general Weibull distribution; the velocity distribution is then no longer Gaussian. The inference methods developed here readily accommodate this case too.

\subsection{Dependent speeds}
\label{rice}
While speeds in discrete-time step-and-turn models are most commonly taken to be independent, perhaps conditional on some behavioural process, it is useful to have the additional flexibility of possible dependence between successive speeds. In the case of Rayleigh speeds---i.e.\ Gaussian velocities---this can be achieved by taking $s_{i},s_{i+1}$ to have a correlated bivariate Rayleigh distribution \citep{Mallik} with equal Rayleigh marginals with scale parameter $\sigma$ and a non-negative correlation parameter $\rho$. 
Then the corresponding conditional distribution is the Rice distribution with scale parameter $\sqrt{\sigma^2(1-\rho^2)}$ and non-centrality parameter $s_i\rho$. For the explicit densities, see the Appendix.

\subsection{Observations and errors}
\label{error}
In the current paper, I make no assumptions about the timing of observations; I assume that only locations (and not velocities) are observed, with i.i.d.\ circular Gaussian error having covariance $\errorSD^2 I$. Incorporating dependence in the errors would be straightforward. I will write $\xx^\text{obs}_k, k=0,\ldots,n$ for the $k$th observation and $t^\text{obs}_k$ for the time at which it is made.
Of course, if errors are large, then the kind of reconstruction undertaken here is likely to be futile. A diffusion approximation is then a possible way forward. Alternatively, the ABC method of \citet{RuizSuarez} could perhaps be extended to the case with error.
On the other hand, if errors are zero, that imposes additional `hard' constraints on the trajectories; rather than address those explicitly, I will show in \S\ref{noerror} that the case of errors with fixed small variance is a suitable substitute. Nevertheless, the error-free case is conceptual useful as a motivation for some details of the methodology.

\section{Inference}
\label{infer}

Exact inference about a velocity-jump process requires sampling possible complete trajectories---times of turns, and the corresponding turning angles and speeds---conditional on a set of observations. Given a full trajectory, inference about the parameters is completely straightforward. 
Here I describe the algorithm for sampling a part of a trajectory, conditional on the model and parameters and on the remainder of the trajectory. 

\subsection{Preliminaries}
\label{prelim}

One useful concept is the \emph{tally}: the sequence of numbers of events in the intervals formed by the observations. 
Thinking in terms of the tally helps to motivate, and describe, some of the ways of updating parts of the trajectory.
\label{collinearity}
Another useful concept is what I will call `collinearity', referring from now on to sequences of observations that are collinear in each spatial dimension, and so not just on a line in the plane, but timed to match a constant velocity too.

While the main trajectory updates modify only a part of the trajectory, it is necessary to find an initial trajectory spanning the times of the whole dataset, $[t^\text{obs}_0,t^\text{obs}_n]$. Using the above ideas, we can do so 
efficiently, so that the number of iterations to reach a reasonable trajectory is not excessive. 

Firstly, by thinking about the error-free case, we can identify potential tallies that will allow segments of path passing through all observations (without permitting turns \emph{at} observations) regardless of any collinearity. 
More precisely, a `minimal' tally is one which includes just enough turns to fit all observations exactly, with the fewest turns possible. 
In general there will not be a unique minimal reconstruction or tally, even in the error-free case. 
Given observations 
$\xx^\text{obs}_0,\ldots, \xx^\text{obs}_n$, a minimal tally consists of $k+1$ entries of 0 and $k$ entries of 2, for some $k\in\{0,\ldots,\left\lceil n/2\right\rceil\}$, arranged alternately, with any remaining entries being 1.
Of course, in the presence of observation error, a path with \emph{any} tally has some chance of generating the observed data, but still the idea of a minimal tally is helpful in thinking about approximation. 

In addition, if there \emph{is} collinearity, the number of turns can be reduced further. Where there are three or more consecutive collinear points, a trajectory exists that fits them exactly while including no turns inbetween them. In addition, if there are adjacent runs of three or more collinear points, with no observations between them, it may be possible to have just a single turn linking them; whether this is possible depends on the locations observed, not just the tally. Again, exact collinearity occurs only in the error-free case (or with probability zero otherwise), but the idea guides the approach when observation error is present.

A possible approach to initialising the trajectory is therefore: 
check the data for approximate collinearity, to within some threshold; 
within any collinear sequences, set the tally to zero; 
between any pair of collinear sequences, choose a tally completely at random from those satisfying the definition of `minimal';
generate independent uniform switching times in each interval defined by the data, according to the tally;
calculate velocities so that the trajectory passes through the observations.
This can be refined by checking for cases where the particular configuration of observations permits an even lower tally, as in the main updates; see \S\ref{special} for details.

\subsection{Trajectory updates}
\label{traj_updates}

For a partial trajectory update, we select beginning and end times, $\tB$ and $\tE$, from the current times of events, or the very beginning or end of the dataset. 
That is, if there are $N$ turn events in the current reconstructed trajectory on $[t^\text{obs}_0,t^\text{obs}_n]$, then $\tB \in \{t^\text{obs}_0, t_1,\ldots,t_N\}$ and 
$\tE \in \{t_1,\ldots,t_N,t^\text{obs}_n\}$, with $\tB<\tE$. 
Interpreting $t_0$ as $t^\text{obs}_0$ and $t_{N+1}$ as $t^\text{obs}_n$, $B$ and $E$ take values in $0,\ldots,N$ and $1,\ldots,N+1$ respectively.
The choice of the joint distribution from which $B$ and $E$ 
are selected will greatly affect the efficiency of the algorithm. If the number of existing turns $L$ between $\tB$ and $\tE$, taking values in $0,\ldots,N$, is too large, then the rejection rate will be too high, while if $L$ is too small then the mixing over trajectories will be too slow. In practice, the joint distribution of $\tB$ and $\tE$ is defined by specifying a distribution for $L$, and then sampling $(\tB, \tE)$ uniformly from values consistent with that value. Specific choices are addressed in \S\ref{implement}.

We then construct a 
proposal for that part of the trajectory, and carry out a Metropolis-Hastings accept/reject update. 
Most of the update types detailed below are reversible jump updates \citep{green}, since they can change the number of turns i.e.\ the dimension of the specification of the path. However, because of the independence in the proposals for the accompanying velocities, accommodating this is straightforward; see \S\ref{HR}.

Let $\T$ be the existing turns inside the interval $(\tB, \tE)$, and let $\M$ denote the tally based on the current reconstructed trajectory, also restricted to $(\tB, \tE)$.
Also let $\xx_B, \xx_E$ be the locations $\xx(\tB)$ and $\xx(\tE)$ respectively, let $\vv_B, \vv_E$ be the velocities up to $\tB$ and starting at $\tE$ respectively, and let $\T^\text{obs}$ and  $\XX^\text{obs}$ be the times and values of the interior observations.
Let $\VV$ be the velocities 
{starting} at times $\T$ \emph{and} at time $\tE$.
Let $\XX$ be the locations at turn times $\T$. 
Dependence on current parameter values is suppressed in describing the trajectory updates, for simplicity.

If $\tB>t^\text{obs}_0$ and $\tE<t^\text{obs}_n$, we regard $\xx_B$ and $\xx_E$ as fixed, and consider resampling the turning points and velocities between them. 
If the update starts at the very beginning or end, this is not quite true due to the observation error, and $\xx_B$ or $\xx_E$ itself must be allowed to vary---details are straightforward but omitted here for brevity.

We use several ways of constructing the proposal; each defines a \emph{type} of update that in itself satisfies detailed balance. In each case, a key stage is how the tally is proposed.

\subsection{Tally proposals}
\label{tally_proposals}

We propose a new tally $\M'$ over the interval $(\tB, \tE)$, 
necessarily of the same length as $\M$, using one of the following types of step and implicitly defining a distribution $q_\M(\M'|\M)$.

\tallyprop{Fixed}
The tally is unchanged; $\M' = \M$. 

\tallyprop{Uniform}
The total number of events in the update is fixed, so $\sum_j m'_j = \sum_j m_j$, but the tally is generated as for a Poisson process conditioned on that total, so that $\M'$ 
has a multinomial distribution, with probabilities proportional to the interval lengths, given by $(t^\text{obs}_{i+1}-t^\text{obs}_i)/(\tE-\tB)$, or at the ends of the interval by $(t^\text{obs}_{i}-\tB)/(\tE-\tB)$ or $(\tE-t^\text{obs}_i)/(\tE-\tB)$.

\tallyprop{Poisson} 
Given the current rate $\lambda$, the tally $\M'$ is proposed as it would be for events of a Poisson process of that rate over the interval being updated; that is, the elements of $\M'$ are simply independent Poisson random variables, with means of the form $(t^\text{obs}_{i+1}-t^\text{obs}_i)\lambda$, or at the ends of the interval of the form $(t^\text{obs}_{i}-\tB)\lambda$ or $(\tE-t^\text{obs}_i)\lambda$.

\tallyprop{Random walk}
The elements of the tally are changed independently according to a random walk; that is, if a current element of the tally is $m_j$, then the corresponding proposed value $m'_j$ is given by 
\begin{align*}
m_j+1 &\text{ with probability } p_\text{up},\\
m_j-1 &\text{ with probability } p_\text{down},\\
m_j &\text{ with probability } 1-p_\text{up}-p_\text{down}.
\end{align*}

\subsection{Event time proposals}

Given the proposed tally $\M'$, new times of turns $\T'$ on $(\tB, \tE)$ are proposed according to some process $q_\T(\T'|\M')$. 
For a `Fixed' proposal, with some probability all the event times are left unchanged. Otherwise, and in all other cases, the simplest way to propose event times is independently and uniformly, corresponding to the Poisson process of events conditioned on the proposed tally. This distribution is used in most updates.
However, there are cases where such proposals are very unlikely to succeed, because the time of a particular event is essentially determined by the observed locations, to within some margin related to observation error. A more constructive approach to proposing event times is likely to be necessary in these special cases. 

\subsubsection{Constructing event times}
\label{special}
Consider the simplest such case, where the proposed tally is $\ldots,0,1,0,\ldots$ over the intervals defined by observation times $\ldots,t^\text{obs}_{k-1},t^\text{obs}_{k},t^\text{obs}_{k+1},t^\text{obs}_{k+2},\ldots$ for some $k$. There is a single proposed event in $(t^\text{obs}_{k},t^\text{obs}_{k+1})$; neglecting observation error, the velocities before and after it are determined by the locations observed at times $t^\text{obs}_{k-1},t^\text{obs}_{k}$ and $t^\text{obs}_{k+1},t^\text{obs}_{k+2}$ respectively. Depending on those velocities, there may be a time $t^*$ at which an event can `join' those two parts of the trajectory. If so, then the time $t$ of the proposed event is sampled from the uniform distribution on the interval $(t^*-\tau,t^*+\tau)$ for some small $\tau$, the interval being appropriately adjusted if $t^*$ is close to $t^\text{obs}_{k}$ or $t^\text{obs}_{k+1}$; otherwise $t$ is sampled from the uniform distribution on $(t^\text{obs}_{k},t^\text{obs}_{k+1})$.

Similar ideas can be applied in other kinds of tally, but 
this constructive approach results in a rather complex calculation of the proposal density---particularly for the reversed proposal, needed for the Hastings ratio---so it is used only sparingly; see \S\ref{implement}.

\subsection{Velocity proposals}
\label{velocities}

Given a set of proposed turning times $\T'$, we need to construct a proposal distribution for the 
velocities $\VV'$. Note that $\T'$ is of length  $L' = \sum_j m'_j$, and $\VV'$ is of length $L'+2$, since it is defined (\S\ref{traj_updates}) to include $\vv_E$, enabling dependence between velocities to be correctly incorporated in the proposals, as well as the velocities on the $L'+1$ subintervals defined by $\T'$.
For a `Fixed' proposal in which the event times are retained, we can use random walk Metropolis-Hastings proposals for the first $L'+1$ velocities, and leave $\vv_E$ unchanged. The acceptance/rejection step is then straightforward (details omitted). In all other cases, we use a Gaussian proposal that is independent of the current velocities $\VV$, constructed in two stages. 

Firstly, we define an `unconditional' or `forward' process, a joint Gaussian distribution for the velocities $\VV'$ given $\T'$ that ignores the data.
This may be very simple, 
or attempt to mimic the actual movement model, but regardless 
it is written as a joint distribution for $\VV'$ (even though $\vv_E$ is fixed) conditional on $\vv_B$. While a Gaussian distribution cannot exactly reproduce the joint distributions of velocities in the model, it can match or approximate the marginal distribution and the expected autocorrelation between velocities. 
Given that the actual process of velocities in the model is Markovian, a natural class of such joint Gaussian distributions is to represent the speeds in each co-ordinate direction as $L'+2$ steps of a stationary first-order auto-regressive process, starting from the appropriate component of $\vv_B$, with mean 0 and variance $\sigma^2$ (matching the actual velocity-jump model), and auto-correlation parameter $\gamma$ which can be chosen freely.
That is, writing each $\vv$ as $(u,v)$, this `pre-proposal' $\tilde{q}_{\VV}(\VV'|\T',\vv_B)$ is given by  
\begin{align*}
(u_{B+1}',\ldots,u_{B+L'+2}')^\transpose
\sim
\dist{N}(u_B\ones, \sigma^2C)
\end{align*}
where \ones\ is a vector of 1s and 
\begin{align*}
C = 
\begin{pmatrix}
1 & \gamma & \gamma^2 & \cdots& \gamma^{L'+1}\\
\gamma & 1 & \gamma&\cdots&\gamma^{L'} \\
\vdots & &\ddots & & \vdots\\
 \gamma^{L'+1} & \gamma^{L'}& \gamma^{L'-1}  & \cdots & 1
\end{pmatrix},
\end{align*}
and similarly for $v_{B+1}',\ldots,v_{B+L'+2}'$.

Since $\xx_B$ is fixed, and locations are linear combinations of times and velocities, that also defines a Gaussian distribution for the locations of the turns and hence, 
since locations at intermediate times are also linear combinations, we have a joint Gaussian distribution with $\XX^\text{obs}$ and $\xx_E$ too, where $\XX^\text{obs} = \{\xx^\text{obs}_j: \tB<t^\text{obs}_j<\tE$\}. 
Writing each $\xx$ as $(x,y)$, for each $k$ such that $t_{B+i}' < t_k^\text{obs} < t_{B+i+1}'$  we have 
\begin{align*}
y_k^\text{obs} &= y_{B+i-1}' + ( t_k^\text{obs}-t_{B+i}')u_{B+i}' + \epsilon_k,
\end{align*}
where
\begin{align*}
y_{B+i}' &= y_{B} + \sum_{j=0}^{i-1} (t_{B+j+1}'-t_{B+j}')u_{B+j+1}'\\
\epsilon_k &\sim \dist{N}(0,\errorSD^2),
\end{align*}
and 
\begin{align*}
y_{E}' &= y_{B} + \sum_{j=0}^{L'} (t_{B+j+1}'-t_{B+j}')u_{B+j+1}',
\end{align*}
and similarly for  $x_{B+1}',\ldots,x_{B+L'+2}'$.
These linear relationships give a joint distribution $\tilde{q}_{\VV,\XX^\text{obs},\xx_E}(\VV',\XX^\text{obs},\xx_E'|\T',\xx_B,\vv_B)$ that takes into account the possible observation errors as well as the expected values obtained by interpolation between the turns.

Secondly, we condition this joint distribution on the observations during the interval, $\XX^\text{obs}$, and on the location and velocity at the end, $\xx_E$ and $\vv_E$, which are being treated as fixed, using the standard form for conditioning of a multivariate Gaussian written in partitioned form. This conditioned joint distribution is then used as the actual proposal for the velocities. For convenience, write $\YY$ for these additional quantities we condition on, so $\YY = (\XX^\text{obs}, \xx_E, \vv_E)$. 
Note that $\VV$ and $\YY$ overlap; $v_E$ is an element of both. We denote the actual proposal distribution by $q_{\VV}(\VV'|\T',\xx_B,\vv_B,\YY)$. 

This conditioning strategy results in a somewhat Gibbs-like proposal, in that it attempts to match the data rather than staying close to the current reconstruction for this part of the trajectory. In general the unconditional process of velocities is not a precise representation of the autocorrelation imposed by the velocity-jump model, but the conditioning does mean that the paths proposed will generally fit the data well. 

\subsection{Acceptance or rejection}
\label{HR}

The final stage is to calculate the appropriate acceptance probability. Excluding the special cases mentioned in Section \ref{special}, the joint proposal is 
\begin{align*}
q(\T',\VV'|\xx_B,\vv_B,\YY,\M)
&= q_{\M}(\M'|\M)q_{\T}(\T'|\M')q_{\VV}(\VV'|\T',\xx_B,\vv_B,\YY).
\end{align*}
Note that this is in part an `independence' proposal, in that it does not depend on $\VV$.
The relevant term from the posterior distribution can be written as
\begin{align*}
p(\T,\VV|\xx_B,\vv_B,\YY) 
&= p(\T,\VV|\xx_B,\vv_B)p(\YY|\T,\VV,\xx_B,\vv_B)/p(\YY|\xx_B,\vv_B)\\
&= p(\T)p(\VV|\T,\vv_B)p(\YY|\T,\VV,\xx_B)/p(\YY|\xx_B,\vv_B),
\end{align*}
where
$p(\T)$ is the Poisson process model for actual events,
$p(\VV|\T)$ is the rest of the actual movement model, based on the distribution of steps and turns and the Jacobian term to transform them into Cartesian velocity space,
and
$p(\YY|\T,\VV)$ is the observation model---i.e.\ Gaussian errors on the interpolated locations---along with a term that is 1 for the probability of $\vv_E$ given $\VV$. 
The acceptance probability is then $\min\{1,h(\T',\VV'|\T,\VV)\}$, where the Hastings ratio $h(\T',\VV'|\T,\VV)$ is given by
\[
\frac{p(\T',\VV'|\xx_B,\vv_B,\YY)}{p(\T,\VV|\xx_B,\vv_B,\YY)} \cdot
\frac{q(\T,\VV|\xx_B,\vv_B,\YY,\M')}{q(\T',\VV'|\xx_B,\vv_B,\YY,\M)} \cdot
\left|\frac{\partial(\T',\VV',\T,\VV)}{\partial(\T,\VV,\T',\VV')}\right|
\]
and the final term is the Jacobian required because of the (possible) dimension change in the trajectory update. Since $\T',\VV'$ do not depend on $\T,\VV$, except through the term $q_{\M}(\M'|\M)$, we have 
\[
\left|\frac{\partial(\T',\VV',\T,\VV)}{\partial(\T,\VV,\T',\VV')}\right| 
= \begin{vmatrix}\mathbf{0} &I_{3L'+4}\\ I_{3L+4}&\mathbf{0} \end{vmatrix}
= 1,
\]
where 
$I_d$ denotes the identity matrix of dimension $d$ and $\mathbf{0}$ denotes a matrix of zeros of the appropriate dimension; the dimensions $3L+4, 3L'+4$ come from the lengths of $\T,\VV,\T',\VV'$ and $\VV,\VV'$ being in 2-dimensional velocity space. 
Because of the way the proposals are tailored to the model and likelihood, the remaining terms in $h(\T',\VV'|\T,\VV)$ can be simplified. We have 
\begin{align*}
&\frac{q(\T,\VV|\xx_B,\vv_B,\YY,\M')}{p(\T,\VV|\xx_B,\vv_B,\YY)} \\
&=
\frac
{q_{\M}(\M|\M')q_{\T}(\T|\M)q_{\VV}(\VV|\T,\xx_B,\vv_B,\YY)}
{p(\T)p(\VV|\T,\vv_B)p(\YY|\T,\VV,\xx_B)/p(\YY|\xx_B,\vv_B)}\\
&=
\frac{q_{\M}(\M|\M')q_{\T}(\T|\M)}{p(\T)} \cdot
\frac{\tilde{q}_{\VV}(\VV|\T,\xx_B,\vv_B)p(\YY|\T,\VV,\xx_B,\vv_B)/p(\YY|\T,\xx_B,\vv_B)}
{p(\VV|\T,\vv_B)p(\YY|\T,\VV,\xx_B)/p(\YY|\xx_B,\vv_B)}\\
&= 
\frac{q_{\M}(\M|\M')q_{\T}(\T|\M)}{p(\T)} 
\cdot
\frac{\tilde{q}_{\VV}(\VV|\T,\xx_B,\vv_B)}{p(\VV|\T,\vv_B)}
\cdot
\frac{p(\YY|\xx_B,\vv_B)}{p(\YY|\T,\xx_B,\vv_B)}
\end{align*}
and similarly for ${q(\T',\VV'|\xx_B,\vv_B,\YY,\M)}/{p(\T',\VV'|\xx_B,\vv_B,\YY)}$.
In each case, the first term corrects for the proposal for event times not matching the Poisson process exactly; the second corrects for the pre-proposal for the velocities not matching the step-and-turn velocity model exactly; the third reflects the two-stage proposal of event times and then velocities, and is easily calculated as both numerator and denominator come from Gaussian densities.

\section{Experiments and Analysis}
\label{apply}

\subsection{Modelling choices and implementation} 
\subsubsection{Modelling}
For the examples here, I take all velocity distributions to be circular bivariate Gaussian, meaning that speeds follow a Rayleigh distribution, and correlated speeds can be represented as described in \S\ref{rice}.
As specified in \S\ref{model}, I will take the distribution of the turning angles to be von Mises with mean zero.

\subsubsection{Prior distributions}

The choice of prior for the model parameters is completely flexible, with their sampling being carried out via Metropolis-Hastings updates. The choices for the analyses here are as follows. 

For the prior on the rate parameter of turn events, $\lambda$, the conjugate Gamma prior is rather restrictive in terms of shape; I use a normal distribution truncated below at zero, making it easy to specify a strictly positive mode without the prior density approaching zero for small values of $\lambda$, or a mode at zero (e.g.\ the special case of a half-normal or folded normal distribution).
Similarly, a truncated normal distribution is used for the standard deviation $\errorSD$ of the observation errors 
and the scale parameter $\sigma$ for speeds.

For the concentration parameter of the von Mises turning angle distribution, the conjugate prior is 
the Bessel Exponential distribution \citep{GuttorpLockhart}, 
$p(\kappa) \propto I_0(\kappa)^{-a}\exp(-b\kappa)$,
where $I_0(\cdot)$ is the zeroth order modified Bessel function of the first kind. The hyperparameters used are $a=1.0, b=-0.5$, chosen `by eye' to avoid an implausibly high prior probability of large concentrations.

The correlation parameter $\rho$ between successive speeds is taken to be $\dist{Beta}(\alpha,\beta)$ with hyperparameters $\alpha=1.0,\beta=1.2$. Thus negative correlations are excluded, as is $\rho=1$, but values in $[0,1)$ are permitted, and in particular values of $\rho$ close to 1 are plausible; this is desirable since the natural parameterisation means that the actual correlation between speeds is rather less than the value of $\rho$.

\subsubsection{Implementation details}
\label{implement}

Since mixing on the rate of turns $\lambda$ is the limiting factor in overall mixing, each experiment is based on two runs, with initial values of $\lambda$ given by the maximum and minimum values of rates calculated approximately from the data. These approximations use a range of values for the standard deviation $\errorSD$ of the observation errors, up to a specified limit (essentially the upper bound of an informative prior) and carry out simple estimation of $\lambda$ based on approximate collinearity in the sense of \S\ref{collinearity}. The threshold used in defining approximate collinearity is 0.1 times the mean empirical speed times the median interval between observations.
The initial reconstruction of the trajectory is based on a random minimal tally as described in \S\ref{prelim}, using the same threshold for approximate collinearity.
The initial value for 
$\errorSD$ is set to half the upper bound; the (unitless) concentration $\kappa$ for the von Mises distribution is set to 2.0, and the initial scale parameter $\sigma$ for speeds is just the empirical speed based on the raw observations.

The four proposal types described in \S\ref{tally_proposals} are taken to be equally likely. 
For a `Fixed' proposal, the probability that the event times are resampled is 1/2, and for a `Random walk' proposal, the steps for each element of the tally have a discrete uniform distribution on $\{-1,0,1\}$, i.e. $p_\text{up} = p_\text{down} = 1/3$, with each proposal $m$ bounded below by 0.

The joint distribution for $\tB$ and $\tE$ is constructed by initially sampling $L$ from  
a distribution that depends on the type of tally update, with parameters set on the basis of experimentation with simulated data:

\begin{tabular}{ll}
Fixed (event times unchanged)&$L\sim\dist{Uniform}\{1,\ldots,30\}$;\\
Fixed (event times resampled)&$L\sim\dist{Uniform}\{1,\ldots,7\}$;\\
Uniform&$L\sim\dist{Uniform}\{1,2\}$;\\
Poisson&$L\sim\dist{Binomial}(9,0.2)$;\\
Random walk&$L\sim\dist{Binomial}(9,0.2)$.
\end{tabular}

\noindent
The proposals for velocities, prior to conditioning, are of the auto-correlated form described in \S\ref{velocities}, but with the auto-correlation parameter $\gamma$ set to 0, as experiments suggest that performance is rather insensitive to $\gamma$ provided that it is not too large.

\subsection{Reindeer data}
\label{reindeer}

As an example application to real data, I re-analysed the same set of 50 GPS fixes on a single reindeer at two minute intervals used by \citet{AlisonBAYSM}.

The upper bound on $\errorSD$ (in metres) used in calculating initial values is taken as 25, based conservatively on properties of the GPS system.
The prior distribution for $\errorSD$ (in metres) is half-normal with scale parameter 10.0.
The prior distribution for $\lambda$ (in turns/second) is half-normal with scale parameter 1/300.

The prior distribution for speed parameter $\sigma$ (in metres/second) is normal, truncated below at zero, with mode and scale parameter both 1/2; the mode is taken from the overall point estimate in \citet{Skarin_speed}, adjusted by treating the movement as diffusive on the hourly timescale of their observations, with a high enough spread that it is largely uninformative. 

\subsubsection*{Results}

\begin{figure}
\begin{center}
\includegraphics[scale=0.75]{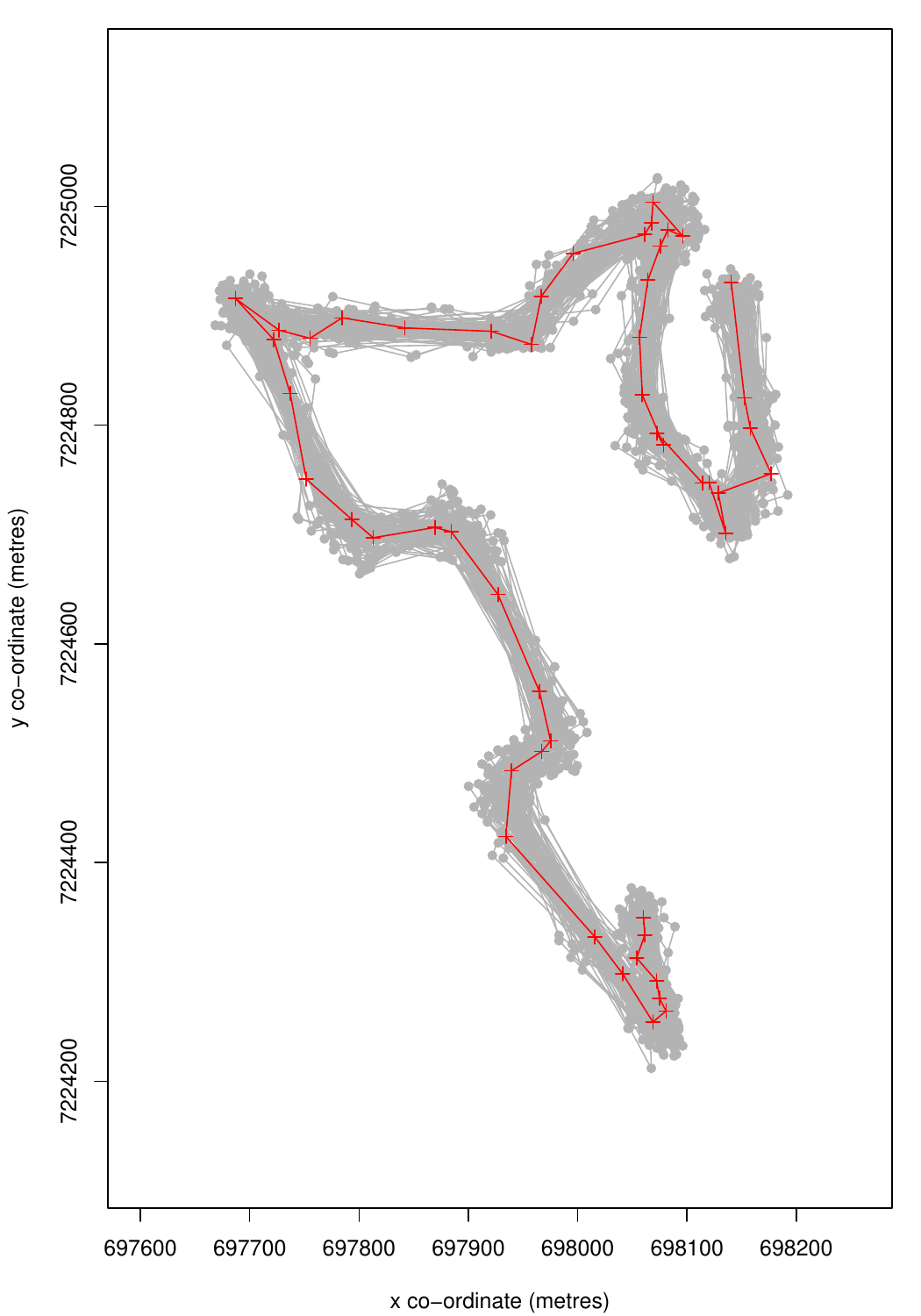}
\caption{100 realisations from the MCMC output based on the reindeer data}
\label{reindeer_dat_2}
\end{center}%
\end{figure}

\begin{figure}
\begin{center}
\includegraphics[scale=0.75]{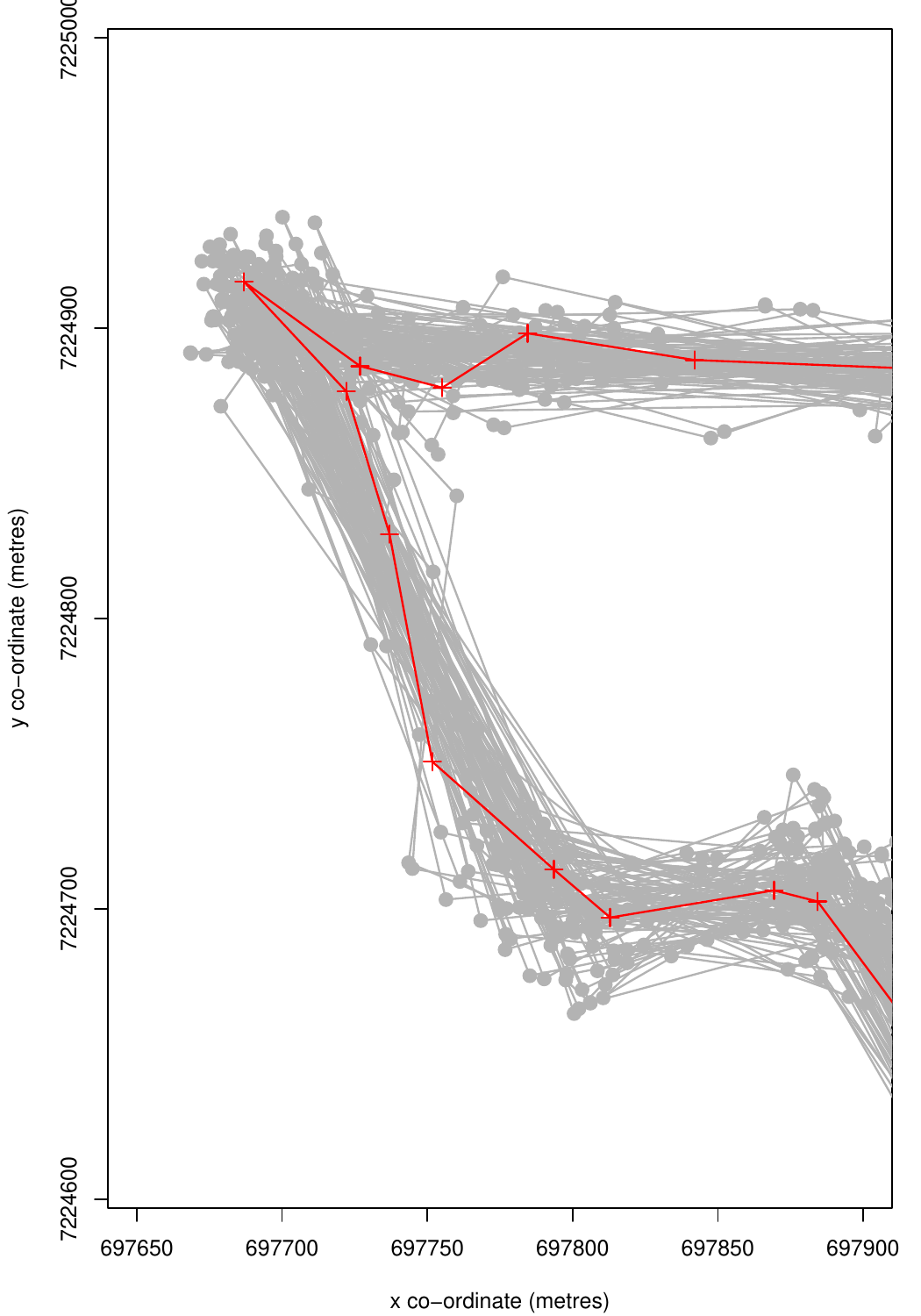}
\caption{A close-up of 100 realisations from the MCMC output based on the reindeer data}
\label{reindeer_dat_4}
\end{center}%
\end{figure}

Based on two runs each of $2\times10^5$ iterations, with the first $5\times10^4$ discarded as burn-in, the posterior distributions are as summarised in Table \ref{results_data}.
Effective sample sizes are 112 for $N$ and over 250 for all the other parameters listed.
Figures \ref{reindeer_dat_2} and \ref{reindeer_dat_4} illustrate these results. 
Fig.\  \ref{reindeer_dat_2} shows the whole dataset analysed (red crosses, linked in time order by red lines) along with a sample of 100 of the trajectories reconstructed by MCMC (grey circles for the reconstructed turns, with grey lines for the constant-velocity segments between them), 
to give an overall impression of the location and the uncertainty in the posterior distribution.
Fig.\ \ref{reindeer_dat_4} shows the same things for a smaller part of the trajectories, concentrating on an interval where the raw data suggest a very sharp turn. 
Figures S1 and S2 (in the Supplementary Information) show the same plots but with a sample of only four trajectories, making it possible to see the details of each individual reconstruction. 

\subsection{Simulated reindeer}
\label{sim}

To explore the performance of the method when the true path and parameters are known, 100 observations were simulated from the model with parameters 
set to the 
posterior means from \S\ref{reindeer}, rounded 
to the Renard series of preferred numbers, $R''(10)$. The actual values used are shown in the table of results below.
To mimic realistic data, error was added as described in \S\ref{error} with $\errorSD = 12$, again based on the estimated value from \S\ref{reindeer}. The analysis was carried out exactly as described for the data in \S\ref{reindeer}.

\subsubsection*{Results}

For the simulated data, posterior distributions for the parameters are summarised in Table \ref{results_sim}.
Effective sample sizes are 98 for $N$ and over 200 for all the other parameters.
Mixing is a little slower than for the real data, since there are more observations. 
Inference is reasonable for all the parameters, with the poorest being the concentration $\kappa$; the true value is comfortably within the equal-tailed 95\% posterior interval, but the posterior median of 0.298 is much lower than the true value of 0.8. This illustrates the difficulty of reconstructing the actual turning angles in the presence of observation error, and when the timing of turns is unknown. For comparison, an analysis based on the true turning angles (possible because this is a simulation) gives a corresponding point estimate of 0.553, so the difference is in part a property of this particular realisation from the model.
Illustrations of the output are given in Figures \ref{reindeer_sim_2} and \ref{reindeer_sim_4} below, and Figures S3 and S4 in the Supplementary Information, along with those for simulated error-free observations.

\subsection{Error-free observations}

\label{noerror}
As mentioned in \S\ref{error}, it is convenient to implement the case without observation error simply by setting $\errorSD$ to some fixed negligible value, rather than separately coding the exact constraints. To illustrate this, the simulated trajectory from \S\ref{sim} is re-analysed here with no errors added and with $\errorSD$ fixed at 0.2 (a factor of 60 less than a realistic value, and so effectively zero; see `Numerical considerations' below) during the inference. 

\subsubsection*{Results}

The results are given in Table \ref{results_reindeer_sim_noerror}.
The effective sample size is 52 for $N$, and at least 300 otherwise.
As expected, the recovery of the true parameter values is consistently more accurate than in the with-error case. The concentration $\kappa$ is apparently still somewhat underestimated, but in fact the posterior median is slightly higher than would be obtained from the exact turning angles in this particular realisation, as described in \S\ref{sim}.

\begin{figure}
\begin{center}
\includegraphics[scale=0.8]{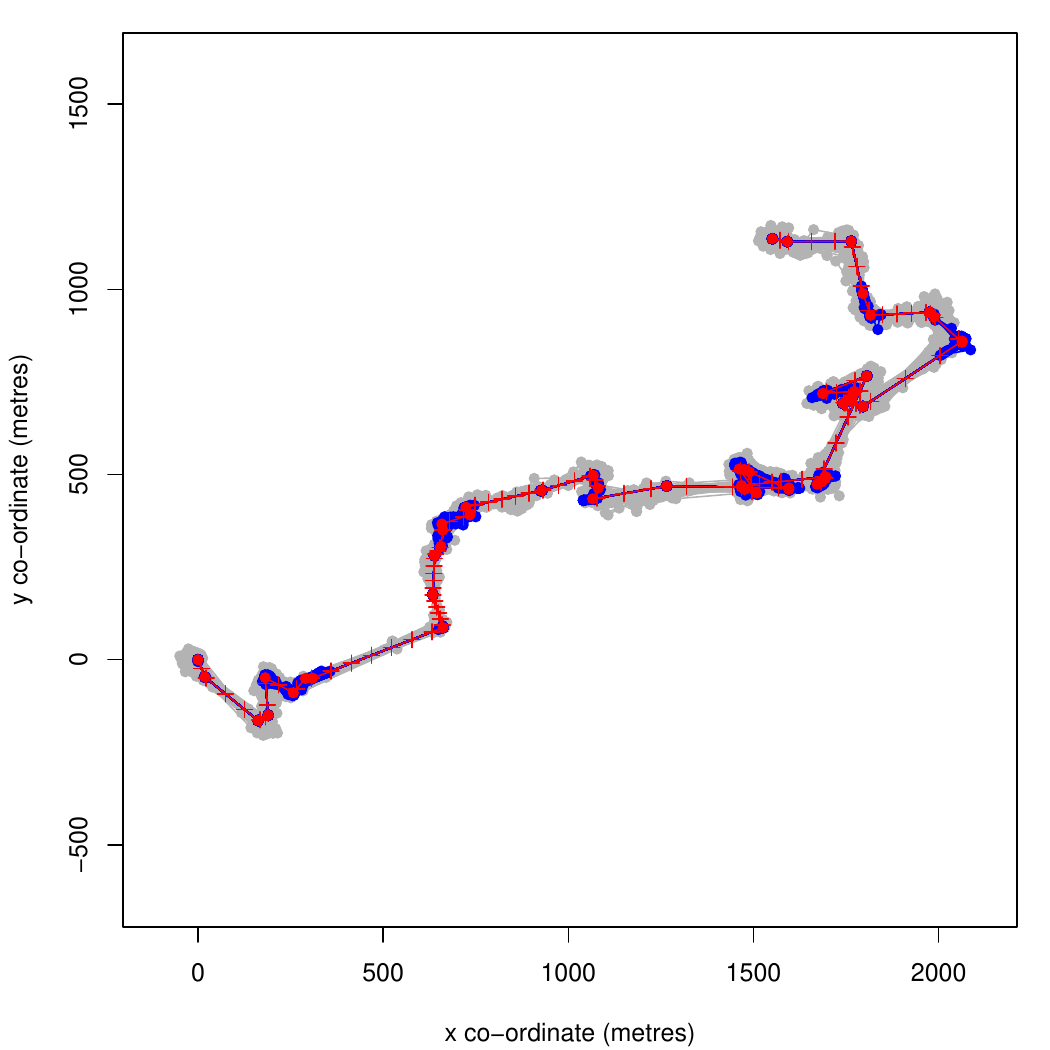}
\caption{100 realisations from the MCMC output based on the simulated data}
\label{reindeer_sim_2}
\end{center}%
\end{figure}

\begin{figure}
\begin{center}
\includegraphics[scale=0.8]{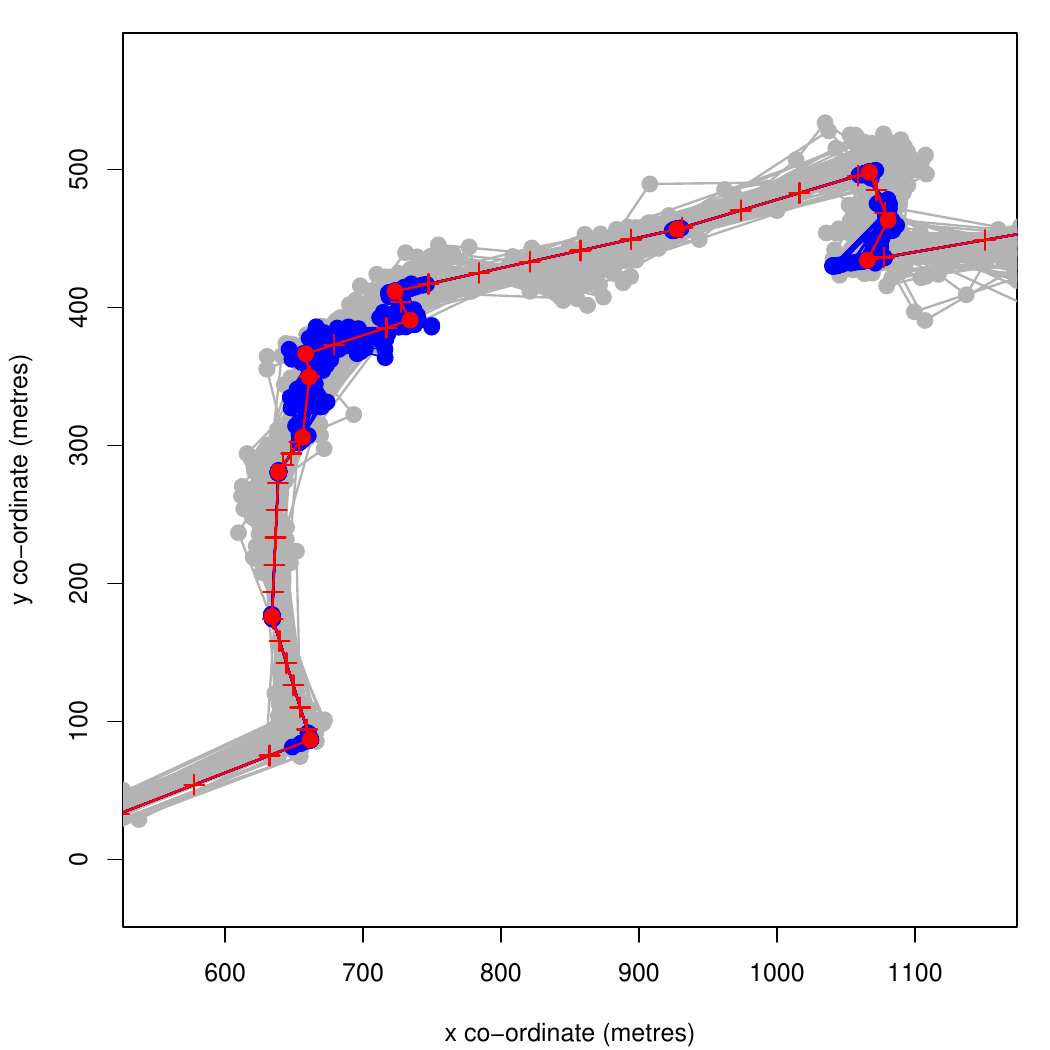}
\caption{A close-up of 100 realisations from the MCMC output based on the simulated data}
\label{reindeer_sim_4}
\end{center}%
\end{figure}

Figures \ref{reindeer_sim_2} and \ref{reindeer_sim_4} show the simulated trajectory and output from both analyses, with and without added measurement error. In each plot, the precise simulated trajectory is shown in red, with circles indicating the turns. Samples from the MCMC output in the absence of measurement error are shown in blue, and those incorporating measurement error in grey. The paths are overlaid in the order grey, blue, red; the limited visibility of the blue reconstructions reflects their similarity to the true paths.
Fig.\ \ref{reindeer_sim_2} shows the simulated set of 100 observations, along with a sample of size 100 from the MCMC analysis. 
There are many times when the trajectories from the analysis without measurement error (in blue) largely coincide with the true trajectory, since the observations suffice to determine the truth with high probability. At other times, when the true path has larger numbers of turns relatively close together, the observations allow real uncertainty about the true path, and the variation in the reconstructions is readily visible. This is particularly clear in the more detailed plot of part of the path in Fig.\ \ref{reindeer_sim_4}. There are two time intervals where successive numbers of turns between observations, i.e.\ the `tally', are high enough to cause such ambiguity in the reconstructions, evident in the variability in the numbers of turns visible (in blue). More generally, the uncertainty in the analysis with measurement error clearly varies between different locations along the path (grey samples).
Figures S3 and S4 in the Supplementary Information show similar plots but with samples of size four from the MCMC analysis, again making it possible to follow each individual reconstruction. 

\subsubsection*{Numerical considerations}
Inference treating observations as coming exactly from the model, with zero error,  would require separate implementation, as mentioned above; but treating them as having very small known error, as here, potentially raises its own issues. Taking the assumed value of $\errorSD$ to be any lower than used here causes numerical issues when calculating eigenvalues very close to zero in the singular Gaussian proposals. More precisely, the observations in all the examples here were in practice scaled so that the typical time interval between them is 1, and the mean speed is also around 1. This helps with general numerical stability and with the tuning of the MCMC proposals. Based on that scaling, an error standard deviation of 0.004, equivalent to $\errorSD = 0.2$ on the scale of the reindeer data as used here, works well, but a standard deviation of 0.002, equivalent to $\errorSD = 0.1$, leads to numerical inaccuracies even when calculating eigenvalues with the high-quality general-purpose code in the R base package \citep{Rmanual}, ultimately based on LAPACK \citep{Lapack}. More carefully customised use of the general-purpose routines could no doubt avoid this issue, but the values used here seem to suffice for present purposes.

\section{Discussion}
\label{discuss}
\subsection{Overview}

In this paper I have shown that it is feasible to carry out fully Bayesian inference for the trajectories and parameters in a piecewise linear or velocity-jump model that is exact in the usual sense for Markov Chain Monte Carlo. This enables a better understanding of likely movement paths actually followed by an animal being tracked. The MCMC output can be interrogated in a wide variety of ways; the brief examples given here show, for example, how reconstructed paths can systematically differ from those obtained from na\"ively interpreting the observed locations through a discrete-time analysis. 

The velocity-jump model is relevant only when the rate of turns, $\lambda$, is not too high; otherwise the trajectory is probably better modelled as a (velocity-based) diffusion as in \citet{johnson2008continuous} and \citet{TheoVelocity}.

\subsection{Approximations}

\label{approx}

Even in the error-free case of the velocity-jump model, we will never \emph{need} more than two turns between observations, and rarely have evidence to be confident of more than two.
The algorithm described in \S\ref{infer} lends itself to a natural approximation, reconstructing the trajectories with as few turning points as possible i.e.\ with a tally that is minimal in the sense of \S\ref{prelim}.
A possible implementation of approximate inference in this sense is as follows. Initially, 
check for approximate collinearity; impose a fixed upper bound on elements of the tally, of 0 within a collinear run of observations, or 1 between adjacent runs if possible (c.f.\ \S\ref{special}), or 2 elsewhere. 
Generate an initial reconstruction which is minimal on the subsequences consisting of 2s in the upper bound, as described in \S\ref{prelim}. 
Whenever the trajectory is updated in a way that affects the tally, require that the new tally, considered globally, is minimal or subminimal.
In practice, initial experiments suggest that any computational gain from this approximation is rather small, without additional customisation of code specifically to exploit it, and therefore the exact approach is more worthwhile.

One alternative---not so far explored---would be to modify the distribution of turning angles to `discourage' small turns. This would effectively favour reconstructions in which events with small turning angles tended to be removed in favour of longer intervals without turns.
A related, more simplistic, approach, which may suffice for relatively high frequency data, would be to opt for full time-discretization, either assuming that turns take place at observation times or approximating the times of turns 
based on the observation interval within which they occur, rather than attempting to reconstruct them exactly. Both these versions are currently being investigated. 
They both have some similarity to a discrete-time step-and-turn model, but potentially interpret near-collinear points as representing an interval without a turn.
The continuous-time formulation would accommodate irregular observations in time---though of course the quality of the approximation might be affected---and would allow a more meaningful interpretation of the parameters.

\subsection{Parton et al.'s continuous-time step-and-turn model}
\label{parton}

The mechanism for proposing velocities in Section \ref{velocities} also permits a different algorithm for inference for the model of \citet{AlisonBAYSM}; see also \citet{AlisonJABES}. As in that paper, we can approximate the process at a grid of times between observations. We hold those times fixed, and update the velocities exactly as described in \S \ref{velocities} and \S\ref{HR} (with some cancellation in the calculation of the Hastings ratio). Potentially this gives more efficient updating of trajectories than the method in \citet{AlisonBAYSM}, since that algorithm was essentially a random walk for bearings combined with a Gibbs-like proposal for speeds conditional on those bearings, whereas the new approach would be Gibbs-like for the velocities directly. This option is implemented in current code, but its performance has not yet been systematically investigated.

\subsection{Extensions}

As described here, the model assumes that the turn rate and movement parameters are constant over time. A natural extension would be to allow multiple behavioural states, as in \citet{Blackwell1997,Blackwell2003} for diffusion processes or \cite{Morales2004} in discrete time, with, say, a continuous-time Markov chain that switches between them and different parameters for each state. Regarding state transitions of the behavioural Markov chain and refreshments of velocity within a state as simply different kinds of events within the same overall structure, broadly the same algorithmic approach could be used for inference. More generally, the rate(s) of events could be allowed to depend on spatial covariates, as in for example \citet{Exact}.

The current formulation also assumes that the distributions of both velocities and errors are Gaussian. Relaxing these assumptions is straightforward; just as currently the unconditioned Gaussian proposals only approximate the correlation structure of the model being fitted, Gaussian proposals could still be used---to permit the crucial conditioning step---as an approximation to actual non-Gaussian distributions for velocities or errors, with the discrepancy handled through the accept/reject step.

\section*{Acknowledgements}
I am grateful to
the Leverhulme Trust for the award of Leverhulme Research Fellowship RF-2020-24, during which 
the work reported here was carried out;
to the Mal\aa\ reindeer herding community, and Prof.\ Anna Skarin of the Swedish University of Agricultural Sciences, for providing the data used in Section \ref{apply}; to
Dr.\ Alison Poulston (n\'{e}e Parton) and Dr.\ Joris Bierkens for 
helpful discussions;
and to Ms.\ Eloise Bray for many helpful comments on an earlier version.

\bibliographystyle{../../../WorkSync/papers/biom}
\bibliography{../../../WorkSync/papers/AStAreview,../../../WorkSync/papers/PGB,../../../FellowSync/literature/fellowship}

\begin{thebibliography}{}

\bibitem[\protect\citeauthoryear{Anderson, Bai, Bischof, Blackford, Demmel,
  Dongarra, Croz, Greenbaum, Hammarling, McKenney, and Sorensen}{Anderson
  et~al.}{1999}]{Lapack}
Anderson, E., Bai, Z., Bischof, C., Blackford, S., Demmel, J., Dongarra, J.,
  Croz, J.~D., Greenbaum, A., Hammarling, S., McKenney, A., and Sorensen, D.
  (1999).
\newblock {\em {LAPACK Users' Guide. Third Edition}}.
\newblock SIAM.

\bibitem[\protect\citeauthoryear{Blackwell}{Blackwell}{1997}]{Blackwell1997}
Blackwell, P.~G. (1997).
\newblock Random diffusion models for animal movement.
\newblock {\em Ecological Modelling} {\bf 100,} 87--102.

\bibitem[\protect\citeauthoryear{Blackwell}{Blackwell}{2003}]{Blackwell2003}
Blackwell, P.~G. (2003).
\newblock Bayesian inference for {Markov} processes with diffusion and discrete
  components.
\newblock {\em Biometrika} {\bf 90,} 613--627.

\bibitem[\protect\citeauthoryear{Blackwell, Niu, Lambert, and
  LaPoint}{Blackwell et~al.}{2016}]{Exact}
Blackwell, P.~G., Niu, M., Lambert, M., and LaPoint, S.~D. (2016).
\newblock Exact {Bayesian} inference for animal movement in continuous time.
\newblock {\em Methods in Ecology and Evolution} {\bf 7,} 184--195.

\bibitem[\protect\citeauthoryear{Davis}{Davis}{1993}]{Davis1993}
Davis, M. H.~A. (1993).
\newblock {\em Markov models and optimization}, volume~49 of {\em Monographs on
  Statistics and Applied Probability}.
\newblock Chapman \& Hall.

\bibitem[\protect\citeauthoryear{Green}{Green}{1995}]{green}
Green, P.~J. (1995).
\newblock Reversible jump {Markov} chain {Monte Carlo} computation and
  {Bayesian} model determination.
\newblock {\em Biometrika} {\bf 82,} 711--732.

\bibitem[\protect\citeauthoryear{Guttorp and Lockhart}{Guttorp and
  Lockhart}{1988}]{GuttorpLockhart}
Guttorp, P. and Lockhart, R.~A. (1988).
\newblock Finding the location of a signal---a {Bayesian} analysis.
\newblock {\em Journal of the American Statistical Association} {\bf 83,}
  322--330.

\bibitem[\protect\citeauthoryear{Johnson, London, Lea, and Durban}{Johnson
  et~al.}{2008a}]{johnson2008}
Johnson, D., London, J., Lea, M., and Durban, J. (2008a).
\newblock Continuous-time correlated random walk model for animal telemetry
  data.
\newblock {\em Ecology} {\bf 89,} 1208--1215.

\bibitem[\protect\citeauthoryear{Johnson, London, Lea, and Durban}{Johnson
  et~al.}{2008b}]{johnson2008continuous}
Johnson, D.~S., London, J.~M., Lea, M.-A., and Durban, J.~W. (2008b).
\newblock Continuous-time correlated random walk model for animal telemetry
  data.
\newblock {\em Ecology} {\bf 89,} 1208--1215.

\bibitem[\protect\citeauthoryear{Mallik}{Mallik}{2003}]{Mallik}
Mallik, R. (2003).
\newblock On multivariate {Rayleigh} and exponential distributions.
\newblock {\em IEEE Transactions on Information Theory} {\bf 49,} 1499--1515.
\newblock IEEE International Symposium on Information Theory, Sorrento, Italy,
  Jun 24-30, 2000.

\bibitem[\protect\citeauthoryear{Michelot and Blackwell}{Michelot and
  Blackwell}{2019}]{TheoVelocity}
Michelot, T. and Blackwell, P.~G. (2019).
\newblock State-switching continuous-time correlated random walks.
\newblock {\em Methods in Ecology and Evolution} {\bf {10},} {637--649}.
\newblock doi: 10.1111/2041-210X.13154.

\bibitem[\protect\citeauthoryear{Morales, Haydon, Frair, Holsiner, and
  Fryxell}{Morales et~al.}{2004}]{Morales2004}
Morales, J., Haydon, D., Frair, J., Holsiner, K., and Fryxell, J. (2004).
\newblock Extracting more out of relocation data: Building movement models as
  mixtures of random walks.
\newblock {\em Ecology} {\bf 85,} 2436--2445.

\bibitem[\protect\citeauthoryear{Othmer, Dunbar, and Alt}{Othmer
  et~al.}{1988}]{Othmer}
Othmer, H.~G., Dunbar, S.~R., and Alt, W. (1988).
\newblock Models of dispersal in biological-systems.
\newblock {\em Journal of Mathematical Biology} {\bf 26,} 263--298.

\bibitem[\protect\citeauthoryear{Parton and Blackwell}{Parton and
  Blackwell}{2017}]{AlisonJABES}
Parton, A. and Blackwell, P.~G. (2017).
\newblock Bayesian inference for multistate `step and turn' animal movement in
  continuous time.
\newblock {\em Journal of Agricultural, Biological and Environmental
  Statistics} {\bf 22,} 373--392.

\bibitem[\protect\citeauthoryear{Parton, Blackwell, and Skarin}{Parton
  et~al.}{2017}]{AlisonBAYSM}
Parton, A., Blackwell, P.~G., and Skarin, A. (2017).
\newblock Bayesian inference for continuous time animal movement based on steps
  and turns.
\newblock In {\em Springer Proceedings in Mathematics \& Statistics {194}:
  Bayesian Statistics in Action}, pages 223--230.

\bibitem[\protect\citeauthoryear{Patterson, Parton, Langrock, Blackwell,
  Thomas, and King}{Patterson et~al.}{2017}]{toby}
Patterson, T.~A., Parton, A., Langrock, R., Blackwell, P.~G., Thomas, L., and
  King, R. (2017).
\newblock Statistical modelling of individual animal movement: an overview of
  key methods and a discussion of practical challenges.
\newblock {\em Advances in Statistical Analysis} {\bf 101,} 399--438.

\bibitem[\protect\citeauthoryear{{R Core Team}}{{R Core Team}}{2021}]{Rmanual}
{R Core Team} (2021).
\newblock {\em R: A Language and Environment for Statistical Computing}.
\newblock R Foundation for Statistical Computing, Vienna, Austria.

\bibitem[\protect\citeauthoryear{Ruiz-Suarez, Leos-Barajas, Alvarez-Castro, and
  Morales}{Ruiz-Suarez et~al.}{2020}]{RuizSuarez}
Ruiz-Suarez, S., Leos-Barajas, V., Alvarez-Castro, I., and Morales, J.~M.
  (2020).
\newblock Using approximate {Bayesian} inference for a ``steps and turns''
  continuous-time random walk observed at regular time intervals.
\newblock {\em PeerJ} {\bf 8,} e8452.

\bibitem[\protect\citeauthoryear{Skarin, Danell, Bergstrom, and Moen}{Skarin
  et~al.}{2010}]{Skarin_speed}
Skarin, A., Danell, O., Bergstrom, R., and Moen, J. (2010).
\newblock Reindeer movement patterns in alpine summer ranges.
\newblock {\em Polar Biology} {\bf 33,} 1263--1275.

\end{thebibliography}

\appendix
\section*{Appendix: Rice and bivariate Rayleigh distributions}
\label{app}
Since the distributions used in Section \ref{rice} are relatively uncommon, I give them in full here.
The correlated bivariate Rayleigh distribution with equal Rayleigh marginals, with scale parameter $\sigma$, and correlation parameter $\rho\geq0$, is
\[
p(s_i,s_{i+1}) 
= \frac{s_i s_{i+1}}{\sigma^2(1-\rho^2)}
\exp\left(-\frac{s_i^2+s_{i+1}^2}{2\sigma^2(1-\rho^2)}\right)
I_0\left(\frac{\rho s_i s_{i+1}}{\sigma^2(1-\rho^2)}\right)
\]
where $I_0(\cdot)$ is the zero-order modified Bessel function of the first kind. The corresponding conditional distribution, the Rice distribution with scale parameter $\sqrt{\sigma^2(1-\rho^2)}$ and non-centrality parameter $s_i\rho$, is
\[
p(s_{i+1}|s_{i}) = \frac{s_{i+1}}{\sigma^2(1-\rho^2)}\exp\left(-\frac{(s_{i+1})^2+(s_i\rho)^2}{2\sigma^2(1-\rho^2)}\right)I_0\left(\frac{s_{i+1}s_{i}\rho}{\sigma^2(1-\rho^2)}\right)\!,~~s_{i+1}\geq0.
\]

\newpage
\section*{Tables and Figures}

\newcommand{\myrows}
{Posterior&\multicolumn{6}{c}{Parameter}\\
quantile&$\lambda$&$\kappa$&$\sigma$&$\rho$&$\errorSD$&$N$\\}

\begin{table}[hbt]
\centering
\begin{tabular}{c|cccccc}
\myrows
   \hline
2.5\% & 0.00202 & 0.093 & 0.234 & 0.013 & 9.56 & 15 \\ 
  25\% & 0.00304 & 0.535 & 0.284 & 0.123 & 11.34 & 19 \\ 
  50\% & 0.00373 & 0.871 & 0.315 & 0.258 & 12.29 & 23 \\ 
  75\% & 0.00448 & 1.225 & 0.348 & 0.417 & 13.33 & 26 \\ 
  97.5\% & 0.00621 & 1.967 & 0.450 & 0.723 & 15.25 & 34 \\ 
  \end{tabular}
\caption{Results for the reindeer data} 
\label{results_data}
\end{table}

\newcommand{\truezero}
{%
\hline
True value&0.004&0.8&0.3&0.25&0.0&48\\
}

\newcommand{\truerows}
{%
\hline
True value&0.004&0.8&0.3&0.25&12.0&48\\
}
\begin{table}[hbt]
\centering
\begin{tabular}{c|cccccc}
\myrows
   \hline
2.5\% & 0.00207 & 0.015 & 0.281 & 0.014 & 10.23 & 29 \\ 
  25\% & 0.00275 & 0.138 & 0.321 & 0.137 & 11.43 & 35 \\ 
  50\% & 0.00315 & 0.298 & 0.347 & 0.289 & 12.14 & 38 \\ 
  75\% & 0.00364 & 0.509 & 0.377 & 0.432 & 12.87 & 42 \\ 
  97.5\% & 0.00470 & 1.015 & 0.450 & 0.680 & 14.38 & 50 \\ 
\truerows
  \end{tabular}
\caption{Results for the simulated reindeer data} 
\label{results_sim}
\end{table}

\begin{table}[hbt]
\centering
\begin{tabular}{c|cccccc}
\myrows
   \hline
2.5\% & 0.00299 & 0.181 & 0.282 & 0.014 & 0.20 & 46 \\ 
  25\% & 0.00365 & 0.452 & 0.312 & 0.133 & 0.20 & 48 \\ 
  50\% & 0.00402 & 0.610 & 0.330 & 0.271 & 0.20 & 49 \\ 
  75\% & 0.00443 & 0.770 & 0.351 & 0.398 & 0.20 & 51 \\ 
  97.5\% & 0.00529 & 1.096 & 0.396 & 0.619 & 0.20 & 53 \\ 
\truezero
  \end{tabular}
\caption{Results for the simulated data without observation error} 
\label{results_reindeer_sim_noerror}
\end{table}

\newpage

\clearpage
\setcounter{figure}{0}
\renewcommand{\thefigure}{S\arabic{figure}}%

\section*{Supplementary Information: additional figures}
\begin{figure}[!h]
\begin{center}
\includegraphics[scale=0.75]{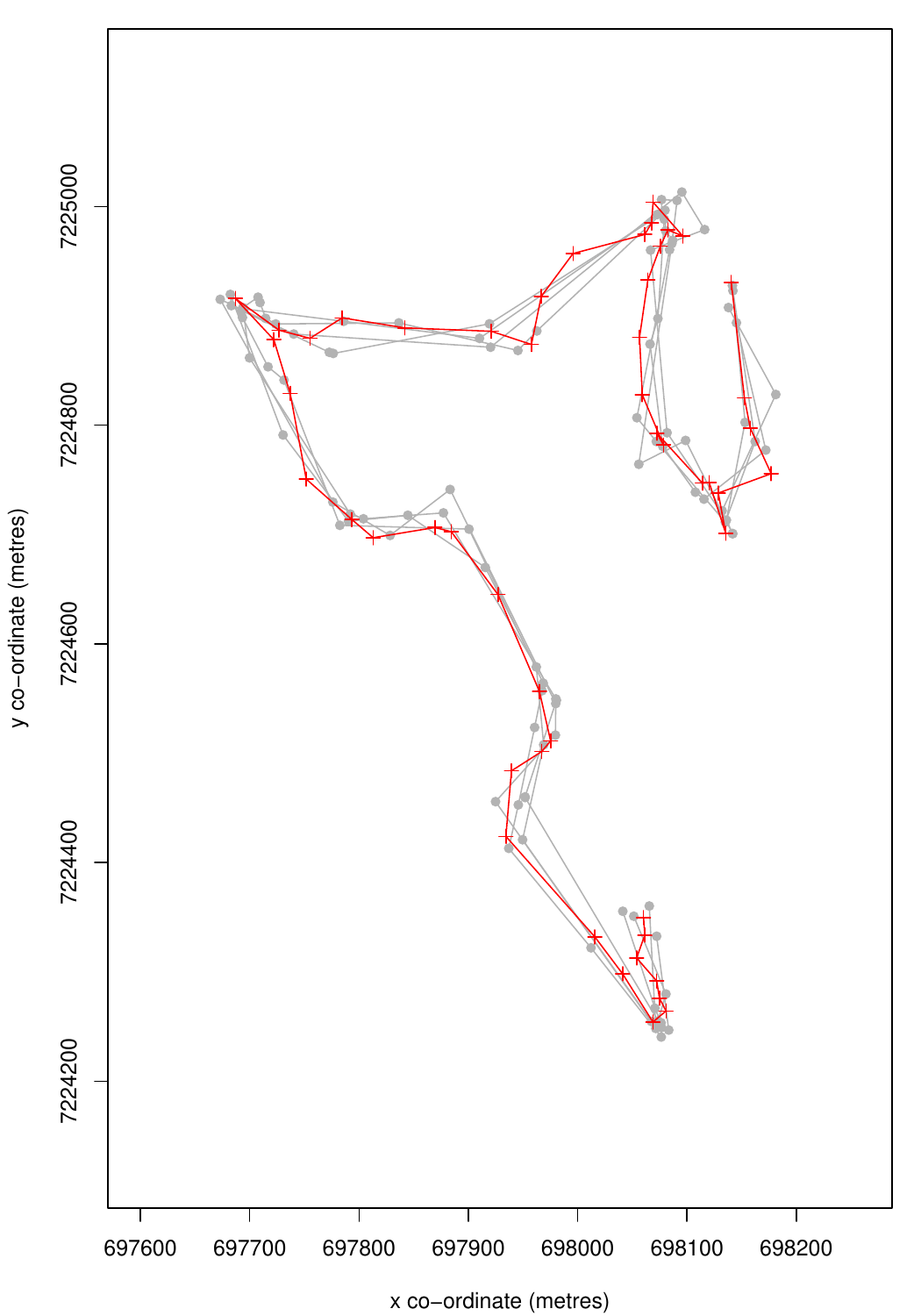}
\caption{Four realisations from the MCMC output based on the reindeer data}
\label{reindeer_dat_1}
\end{center}%
\end{figure}%
\begin{figure}
\begin{center}
\includegraphics[scale=0.75]{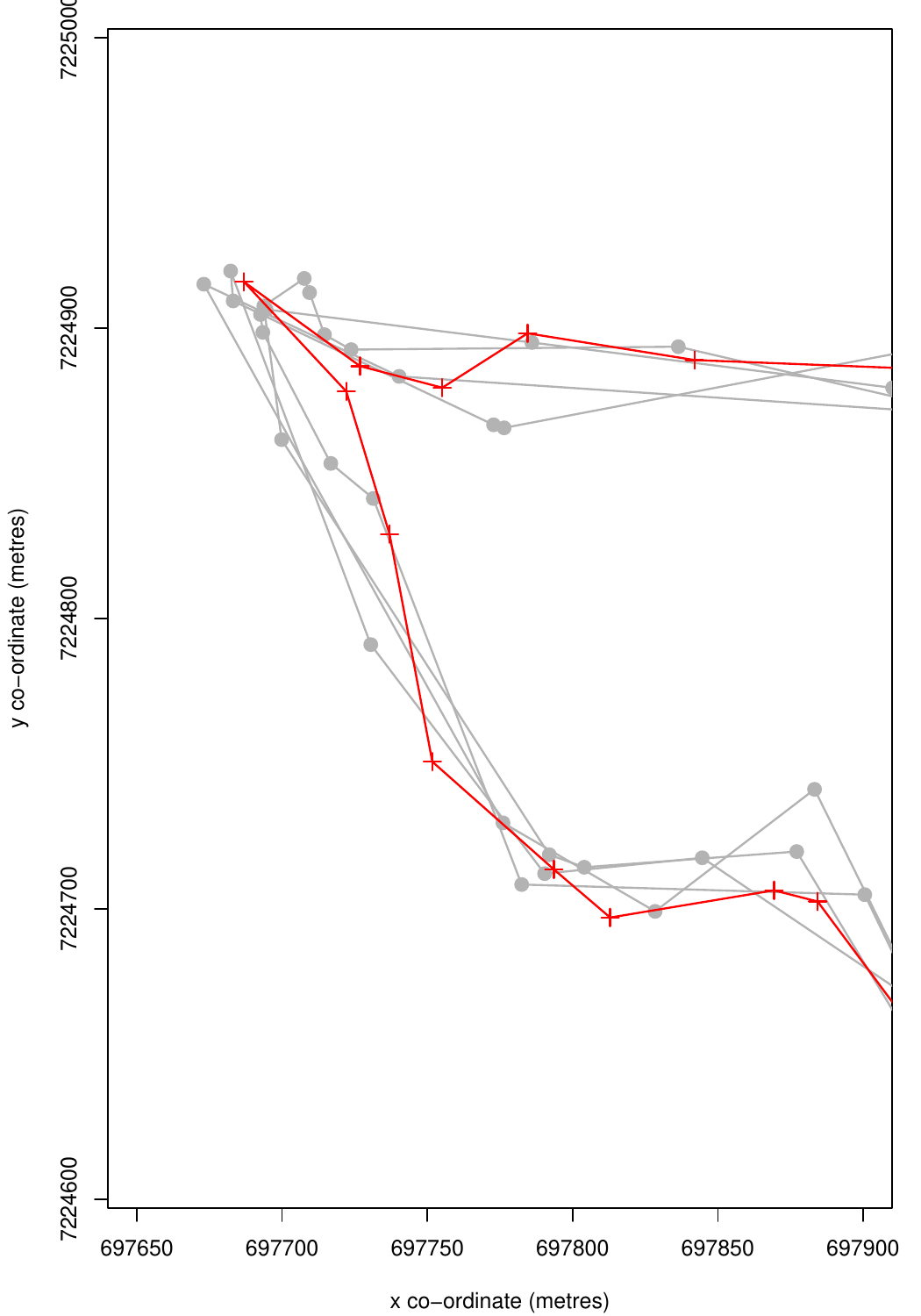}
\caption{A close-up of four realisations from the MCMC output based on the reindeer data}
\label{reindeer_dat_3}
\end{center}%
\end{figure}%
\begin{figure}
\begin{center}
\includegraphics[scale=0.8]{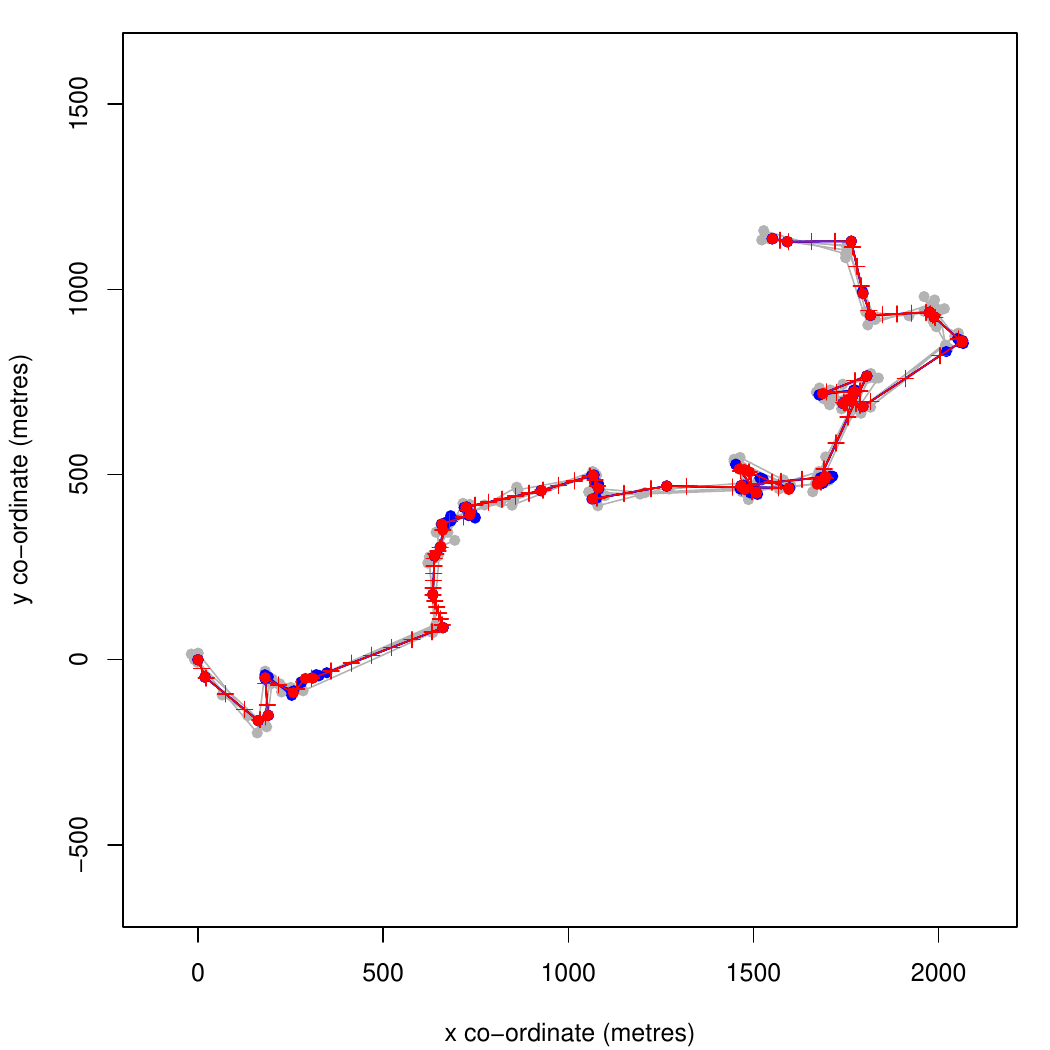}
\caption{Four realisations from the MCMC output based on the simulated data}
\label{reindeer_sim_1}
\end{center}%
\end{figure}%
\begin{figure}
\begin{center}
\includegraphics[scale=0.8]{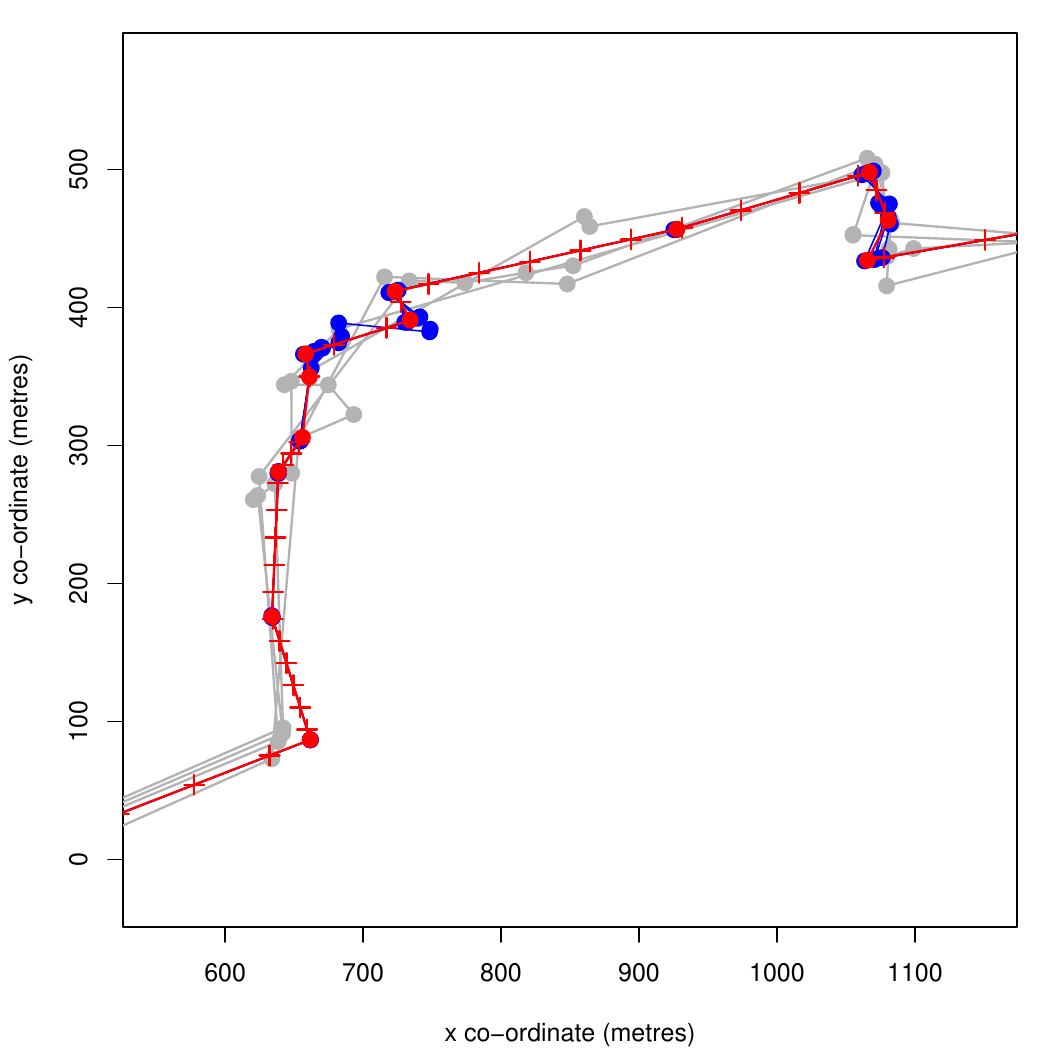}
\caption{A close-up of four realisations from the MCMC output based on the simulated data}
\label{reindeer_sim_3}
\end{center}%
\end{figure}%

\end{document}